\documentclass[sigconf,screen]{acmart}
\usepackage{algorithmic}
\usepackage{graphicx}
\usepackage{textcomp}
\usepackage{xcolor}
\usepackage[ruled]{algorithm2e}
\usepackage{listings}
\usepackage{color}
\usepackage{enumitem}
\usepackage{multirow}
\usepackage{pifont}
\usepackage{subfigure}
\usepackage{verbatim}
\usepackage{caption}
\usepackage{booktabs}
\usepackage{setspace}
\usepackage{arydshln}

\newcommand{\halffigin}{\vspace{-0.04in}}

\newcommand{\figin}{\vspace{-0.08in}}

\definecolor{dkgreen}{rgb}{0,0.6,0}
\definecolor{gray}{rgb}{0.5,0.5,0.5}
\definecolor{mauve}{rgb}{0.58,0,0.82}

\lstdefinelanguage{diff}{
  basicstyle={\footnotesize\ttfamily},
  numberstyle=\footnotesize\color{gray},
  morecomment=[f][\color{red}]{-},  
  morecomment=[f][\color{green!50!black}]{+} 
}

\AtBeginDocument{%
  }

\setcopyright{acmlicensed}
\copyrightyear{2025}
\acmYear{2025}
\acmDOI{XXXXXXX.XXXXXXX}
  
\setcopyright{none}
\renewcommand\footnotetextcopyrightpermission[1]{}
\begin{document}

\title{\textsc{DaiFu}: In-Situ Crash Recovery for Deep Learning Systems}

\author{Zilong He}
\email{hezlong@mail2.sysu.edu.cn}
\affiliation{%
  \institution{School of Computer Science and Engineering, Sun Yat-sen University}
  \country{China}
}

\author{Pengfei Chen}
\authornote{Pengfei Chen is the corresponding author.}
\email{chenpf7@mail.sysu.edu.cn}
\affiliation{%
  \institution{School of Computer Science and Engineering, Sun Yat-sen University}
  \country{China}
}

\author{Hongyu Zhang}
\email{hyzhang@cqu.edu.cn}
\affiliation{%
  \institution{School of Big Data and Software Engineering, Chongqing University}
  \country{China}
}

\author{Xiaoyun Li}
\email{lixy223@mail2.sysu.edu.cn}
\affiliation{%
  \institution{School of Computer Science and Engineering, Sun Yat-sen University}
  \country{China}
}

\author{Guangba Yu}
\email{yugb5@mail2.sysu.edu.cn}
\affiliation{%
  \institution{School of Computer Science and Engineering, Sun Yat-sen University}
  \country{China}
}

\author{Hongyang Chen}
\email{chenhy95@mail2.sysu.edu.cn}
\affiliation{%
  \institution{School of Computer Science and Engineering, Sun Yat-sen University}
  \country{China}
}

\author{Zibin Zheng}
\email{zhzibin@mail.sysu.edu.cn}
\affiliation{%
  \institution{School of Software Engineering, Sun Yat-sen University}
  \country{China}
}

\renewcommand{\shortauthors}{He et al.}
\newcommand{\hy}[1]{{\color{red}[HY: #1]}}

\newcommand{\pf}[1]{{\color{orange}[PF: #1]}}
\newcommand{\zl}[1]{{\color{blue}#1}}

\newcommand{\zlcomment}[1]{{\color{dkgreen}[ZL:#1]}}
\begin{abstract}
Deep learning (DL) systems have been widely adopted in many areas, and are becoming even more popular with the emergence of large language models. However, due to the complex software stacks involved in their development and execution, crashes are unavoidable and common. Crashes severely waste computing resources and hinder development productivity, so efficient crash recovery is crucial. Existing solutions, such as checkpoint-retry, are too heavyweight for fast recovery from crashes caused by minor programming errors or transient runtime errors. Therefore, we present \textsc{DaiFu}, an in-situ recovery framework for DL systems. 
Through a lightweight code transformation to a given DL system, \textsc{DaiFu} augments it to intercept crashes in situ and enables dynamic and instant updates to its program running context (e.g., code, configurations, and other data) for agile crash recovery.
Our evaluation shows that \textsc{DaiFu} helps reduce the restore time for crash recovery, achieving a 1327$\times$ speedup compared with state-of-the-art solutions. Meanwhile, the overhead of \textsc{DaiFu} is negligible (under 0.40\%). We also construct a benchmark spanning 7 distinct crash scenarios in DL systems, and show the effectiveness of \textsc{DaiFu} in diverse situations. 
\end{abstract}




\maketitle

\section{Introduction}




Deep Learning (DL) systems have been extensively developed to support various applications, spanning natural language processing~\cite{DBLP:journals/corr/WuSCLNMKCGMKSJL16}, autonomous vehicles~\cite{DBLP:journals/tits/KuuttiBJBF21}, and intelligent operations~\cite{DBLP:conf/kbse/HeCLYCYL22, DBLP:conf/www/LuoZQW0WLDRLZ21}.
These systems generally work through iterating over data to train deep neural networks, or performing inference using the trained networks. 
Due to the complex software stacks inherent in their development and execution, crashes are inevitably common, leading to substantial computation waste and a significant impediment to development productivity. Therefore, it is crucial to resolve crashes efficiently.

\begin{figure}[tp]
  \centering
  \includegraphics[width=0.9\linewidth]{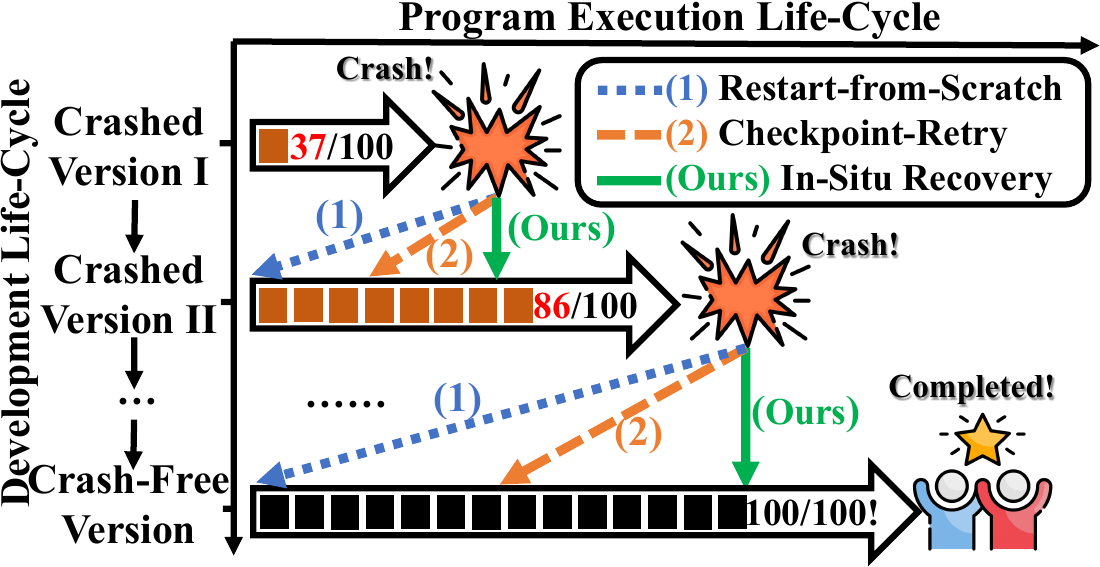}
  \figin
  \caption{Three crash recovery paradigms for DL systems with their recovery processes highlighted by different arrows. Different from prior paradigms, this paper proposes \textit{In-Situ Recovery} to recover crashes, thus contributing to agile development and stable execution of DL systems.}
  \figin
\label{workflow}
\end{figure}


Crash resolution for DL systems includes three processes, i.e., detection, debugging, and recovery. Specifically, (1) Detection~\cite{deng2024minder} denotes a process of recognizing that a crash has occurred. (2) Debugging~\cite{DBLP:conf/icse/WardatCLR22, DBLP:conf/icse/ZhangZMS21, DBLP:conf/icse/CaoLC0TWC22, DBLP:conf/icse/WardatLR21, DBLP:conf/chi/SchoopHH21, DBLP:conf/icse/IslamPNR20, DBLP:conf/icst/KimHJTY23, DBLP:conf/issre/WuSC0Q21, DBLP:journals/tosem/WanLXLHML24} refers to the process of identifying, analyzing, and resolving the root cause that causes a system to crash. (3) Recovery~\cite{DBLP:conf/usenix/KimJLLJ22, DBLP:conf/fast/MohanPC21, DBLP:conf/icml/0036LRJ20, DBLP:conf/nsdi/EisenmanMIMKNSA22, DBLP:conf/sosp/WangJZZFNW23, DBLP:conf/eurosys/GuptaKKVGKRS24} refers to the process of restoring a crashed system to a normal state after the root cause of the crash has been resolved. The restore time is defined as the total time taken to restart and re-run the DL system to the point where it was before the crash (or as close as possible). Prolonged restore time will lead to significant waste of computing resources. Moreover, during the development of DL systems, frequent interruptions by crashes and long waiting time for each restoration can hinder iterative experimentation, and reduce the overall development productivity. \textbf{Therefore, this paper investigates how to reduce the restore time for crash recovery.} 
Existing practices generally take several minutes and even hours to complete restoration~\cite{DBLP:conf/icse/ZhangXZLLY20, DBLP:conf/usenix/JeonVPQXY19, DBLP:conf/nsdi/Hu0WWZC0L0L0024}.
Specifically, these practices fall into two paradigms, i.e., Restart-from-Scratch and Checkpoint-Retry. The \textit{Restart-from-Scratch} paradigm 
restarts the crashed DL system from the beginning of its execution, which generally takes several hours or days for restoration. While the \textit{Checkpoint-Retry} paradigm~\cite{DBLP:conf/usenix/KimJLLJ22, DBLP:conf/fast/MohanPC21, DBLP:conf/icml/0036LRJ20, DBLP:conf/nsdi/EisenmanMIMKNSA22, DBLP:conf/sosp/WangJZZFNW23, DBLP:conf/eurosys/GuptaKKVGKRS24} allows the crashed DL system to retry from one of its previous checkpoints, which generally takes several minutes or hours for restoration. The restore time also increases accordingly as the modern DL models grow larger. Recently, the main optimization direction of state-of-the-art solutions~\cite{DBLP:conf/fast/MohanPC21,DBLP:conf/nsdi/EisenmanMIMKNSA22,  DBLP:conf/sosp/WangJZZFNW23} is to achieve a more frequent checkpointing via asynchronism.

Though prior solutions can reduce the restore time for crash recovery, there is still a complicated trade-off between the achieved restore time, required runtime overhead, and preparation effort. Specifically, developers need to refactor their programs to integrate the corresponding libraries for the interleaving of checkpointing and computations, and despite asynchronism, frequent checkpointing still inevitably introduces extra overhead. The need to balance such trade-offs will distract developers in the early development stages of DL systems. Besides, always requiring a full program restart for crash recovery is too heavyweight, and may be unnecessary for the recovery from crashes caused by minor programming errors or transient runtime errors. Therefore, it is meaningful to provide a new option for DL system crash recovery, which circumvents such complicated trade-offs.



To this end, we explore a new paradigm, i.e., \textit{In-Situ Recovery}, to speed up crash recovery for DL systems. As shown in Fig.~\ref{workflow}, ``in-situ'' means that a crash is intercepted and then recovered with Dynamic Software Updating (DSU)~\cite{DBLP:journals/toplas/HicksN05, DBLP:journals/iet-sen/AhmedLSZ20}. This new paradigm is based on two observations and one key idea. The first observation is that the restore time depends on the freshness of the preserved program context (i.e., the executed code locations and all the variables). The second observation is that the running program context is not corrupted in many crash scenarios, and the affected DL system will throw an interceptable exception before its total crash. Based on these two observations, our key idea is that for crashes whose root cause does not corrupt the running program context, if we intercept the exception and perform crash recovery directly based on the program running context rather than its historical context recorded by checkpoints, the restore time can be greatly reduced. This is what we achieve via in-situ recovery.
However, achieving in-situ recovery for DL systems poses several challenges. 
\textbf{(1) In-situ recovery needs an ultra-fine DSU granularity.} Prior DSU solutions developed for Python online services~\cite{DBLP:conf/usenix/HuangXZ021, DBLP:conf/apsec/TangZ18} cannot be applied to active functions, and thus are not in-situ. 
The entry function (i.e., the main or training function) of a DL system is long-running, and cannot wait for the next call to activate its dynamic updates. That is, the update should be activated immediately regardless of whether the updated function exits or not. Therefore, the first challenge is how can we enable ultra-fine-grained DSU to \textbf{reduce the restore time for crash recovery}? \textbf{(2) In-situ recovery needs a lightweight and convenient DSU trigger.} Achieving low runtime overhead and small preparation effort are also key concerns for a practical in-situ recovery method. For example, hacking into a specific DL framework or even the Python interpreter to achieve some functionalities is not desirable, since this makes the method less general. Therefore, the second challenge is how can we \textbf{reduce the preparation effort and runtime overhead} when enabling the trigger of DSU in the event of program crashes?

In this paper, we propose \textsc{DaiFu}, a framework that vaccinates \underline{D}L systems to enable in-situ recovery ag\underline{ai}nst crash \underline{F}ail\underline{u}res that trigger exceptions before corrupting the program context. Here, ``vaccinate'' means that a DL system is augmented to intercept crashes and allow in-situ updates. 
Specifically, when a crash occurs, \textsc{DaiFu} intercepts it, and interacts with developers to perform crash recovery, and supports statement-level retries, dynamic code execution, instant code updates, and environment changes to aid the recovery process. (1) To \textbf{reduce the restore time for crash recovery}, \textsc{DaiFu} automatically decomposes and reconstructs the entry functions (i.e., the main or training function) in advance to enable their instant and fine-grained updates against crashes. (2) To \textbf{reduce the preparation effort and runtime overhead}, \textsc{DaiFu} is designed to be plug-and-play (usually only two lines of code need to be added, including one line for library import and one line for function decoration), and then \textsc{DaiFu} automates the processes of program running context protection and resume. Moreover, these processes are designed to rely only on lightweight operations, thus achieving low overhead.
Our evaluation demonstrates that \textsc{DaiFu} significantly reduces the restore time for crash recovery, i.e., achieving a speedup of 1327$\times$ compared to existing open-source DL crash recovery solutions. Moreover, the runtime overhead of \textsc{DaiFu} is low (under 0.40\%). Additionally, we establish an executable benchmark comprising 7 distinct crash scenarios to test the feasibility of in-situ recovery in different situations. \textsc{DaiFu} succeeds in enabling the in-situ recovery for 31 out of 32 crash cases, underscoring its usefulness in different situations. 


In general, this paper makes the following contributions.
\begin{itemize}[leftmargin=*]

\item \textbf{Perspective.} To the best of our knowledge, we are the first to introduce and improve DSU to recover crashes of DL systems in situ. This paradigm is substantially different from prior checkpoint-retry based methods, and in many crash scenarios, it can achieve almost instant restoration for crash recovery.

\item \textbf{Framework.} We propose a novel framework \textsc{DaiFu} to implement the in-situ recovery paradigm. With the introduction of program vaccination to prepare for in situ recovery, \textsc{DaiFu} achieves quick crash recovery, while 
incurring low runtime overhead and small preparation effort.
\item \textbf{Benchmark.} We reproduce and extend a benchmark of crash scenarios in DL systems. The benchmark contains reproducible crashes caused by code defects, environment problems, or exceptional data. This benchmark can enable a comprehensive evaluation of related techniques. 
\item \textbf{Implications.} We evaluate \textsc{DaiFu} on the benchmark, demonstrating its quick crash recovery ability, low runtime overhead, and applicability in various crash scenarios. The artifact of \textsc{DaiFu} is available at \url{https://anonymous.4open.science/r/DaiFu}.
\end{itemize}


\section{Background and Motivation}

\subsection{Crash Recovery for DL systems}\label{sec:crash} 
\begin{figure}[tp]
  \centering
  \includegraphics[width=1\linewidth]{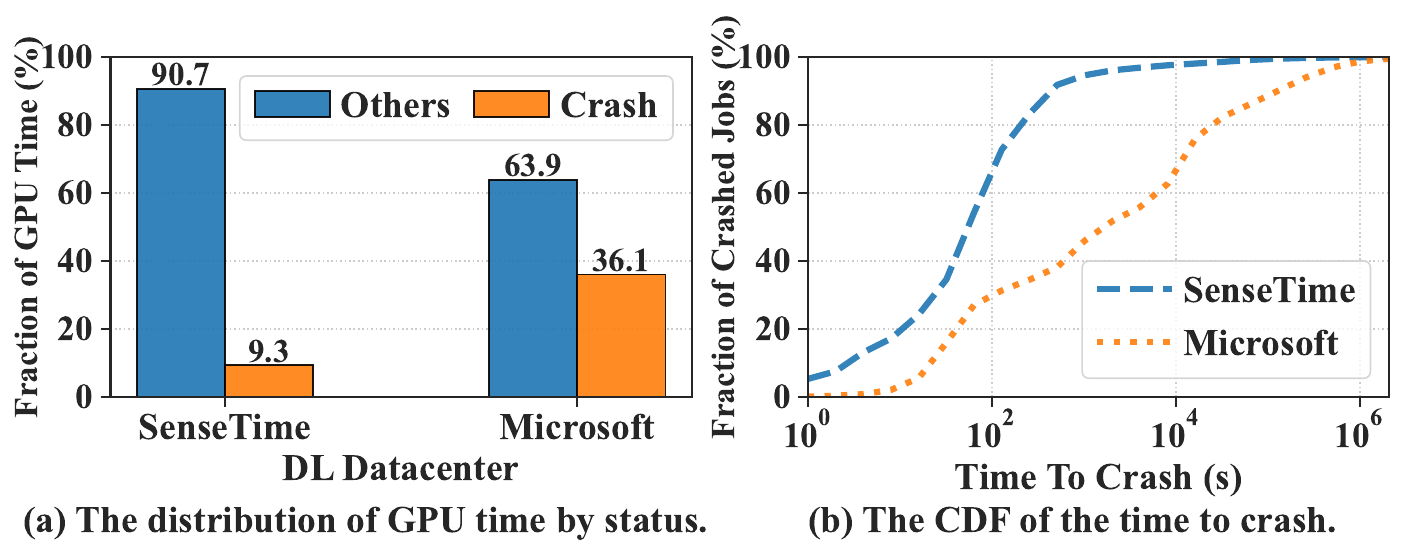}
  \figin
  \figin
  \caption{An analysis of crashes in DL systems from SenseTime Helios~\cite{DBLP:conf/sc/Hu0Y0021} and Microsoft Philly~\cite{DBLP:conf/usenix/JeonVPQXY19}.}
\label{trace_study}
\figin
\end{figure}

Crashes are common in the development and execution of DL systems. They can be caused by many reasons, including code defects, environment problems, and data corruption~\cite{DBLP:conf/icse/ZhangXZLLY20, DBLP:conf/usenix/JeonVPQXY19}. 
Frequent crashes lead to substantial computation waste.
To illustrate, Fig.~\ref{trace_study} presents an analysis of public traces from two DL platforms~\cite{DBLP:conf/sc/Hu0Y0021, DBLP:conf/usenix/JeonVPQXY19}. Based on these traces, we can understand that the effect of crashes in DL systems is non-negligible in practice.
Specifically, in Fig.~\ref{trace_study}(a), we can see that a considerable amount of time in the development of the DL system is affected by program crashes, and from Fig.~\ref{trace_study}(b), we can see that the time to crash can be long (e.g., from $10^4$ to $10^6$s), rendering a restart from scratch expensive and inefficient. Moreover, from the perspective of a DL developer, it is frustrating to see that the work completed by a long-running program needs to be redone due to a crash. Therefore, it is meaningful to reduce the restore time for crash recovery.

\subsection{Dynamic Software Updating}\label{sec:dsu}

Dynamic Software Updating (DSU)~\cite{DBLP:journals/toplas/HicksN05, DBLP:journals/iet-sen/AhmedLSZ20} refers to the process of updating code or data in a running program without requiring it to be restarted. This technique can facilitate continuous operation without downtime. Hence, it is useful in many scenarios~\cite{DBLP:conf/usenix/HuangXZ021, DBLP:conf/apsec/TangZ18, DBLP:conf/icse/ChenYCZY07, DBLP:conf/usenix/HjalmtyssonG98, DBLP:conf/osdi/RommelDFKBMSL20, DBLP:conf/eurosys/NicoaraAR08, DBLP:conf/pldi/SubramanianHM09, DBLP:conf/oopsla/PinaVH14}. For example, developers can use this technique to incrementally upgrade the system without restarting it at each step. Therefore, this paper investigates how to introduce and enhance this technique for the development of DL systems. Next, we will elaborate on its potential benefit, feasibility, and adoption challenges for crash recovery for DL systems.


\textbf{The Potential Benefit of DSU for Crash Recovery.} DSU can allow an almost instant completion of restoration for crash recovery. 
To illustrate, we provide an example to show how DSU can speed up crash recovery for DL systems. Fig.~\ref{example_fault} shows a crash caused by the unexpected data distribution of the input data. Specifically, the labels inputted to the AUC calculation only contain one class, resulting in an exception with ``ValueError: Only one class present in y\_true. ROC AUC score is not defined in that case'' (from Line 15). The effectiveness of existing crash recovery solutions for this crash depends on when a checkpoint is generated. For example, if the checkpointing code is placed at the end of each epoch (e.g., at Line 16 in Fig.~\ref{example_fault}), the restore time for crash recovery is long, since a full training epoch needs to be repeated for a complete recovery. 
The restore time may be reduced depending on the checkpoint frequency, but this requires extra runtime overhead. 
Besides, with only checkpoints and the exception message, it may be difficult for the developer to figure out why the inputted labels contain only one class. However, with in-situ recovery provided by \textsc{DaiFu}, the developer can immediately interact with the program to determine which valid\_file causes the crash, and then he / she can skip the faulty valid\_file with a try-except statement, which is dynamically added to the main function using \textsc{DaiFu}. Then the program can instantly recover from the crash and continue to execute. 


\begin{figure}[tp]
\begin{lstlisting}[language=Python, escapeinside={(*@}{@*)}]
from sklearn.metrics import roc_auc_score
import daifu
...
@daifu.transform() # <- How to integrate DaiFu into a program
def main():
    model, optimizer, args = creat_model() 
    for i in range(args.steps)
        train(model, optimizer) 
        auc_list = []
        for valid_file in file_list:
            valid_data = read_from_file(valid_file)
            x_valid, labels = extract_labels(test_data)
            scores = model.predict(x_valid)
            (*@\color{gray}\# The following code throws Value Error caused by labels with only one class@*)
            (*@\color{red}auc\_list.append(roc\_auc\_score(labels,scores))@*) 
        checkpoint(auc_list, model, optimizer, args)
\end{lstlisting}
\halffigin
\caption{An example of crash caused by the
unexpected data.}
\figin
\label{example_fault}
\end{figure}

\textbf{The Feasibility of DSU for Crash Recovery.} 
It is feasible to use DSU to improve crash recovery for DL systems. (1) \textbf{The Support of the Python Runtime.} The Python interpreter supports DSU through several key features, including its meta-object protocol and dynamic typing, which allow programs to manipulate metadata and variable types at runtime. The only problem is that Python does not support DSU to active functions, which will be discussed in the next paragraph. \textbf{(2) The Existence of In-Situ Recoverable Crashes.} There exist many crash scenarios that can be recovered with DSU. Specifically, many crashes~\cite{DBLP:conf/icse/ZhangXZLLY20, DBLP:conf/usenix/JeonVPQXY19, DBLP:conf/nsdi/Hu0WWZC0L0L0024} manifest as exceptions (e.g., a File Not Found Error) thrown across the function call stack, and will finally reach the user-perceived functions and only crash the program when they are uncaught. Therefore, they can be intercepted by an exception handling statement instrumented into user programs. Besides, a majority of crashes are actually caused by programming errors introduced by developers. For example, a total of 2294 crashes~\cite{DBLP:conf/nsdi/Hu0WWZC0L0L0024} in the development of InternLM~\cite{DBLP:journals/corr/abs-2403-17297}, a large language model, are caused by programming errors of developers, and these crashes are often fixed via revising configurations or codes. Intercepting such crashes and recovering them in situ can greatly accelerate the development progress.

\textbf{The Adoption Challenges of DSU for Crash Recovery.} 
Directly updating an active function is not allowed when applying DSU to Python programs.
This is because the Python interpreter implements code objects on the evaluation stack as read-only. As a result, only when a function totally terminates, the updates to its code are activated. Although some prior work~\cite{DBLP:conf/usenix/HuangXZ021, DBLP:conf/apsec/TangZ18} has investigated DSU for Python online services, their supported DSU is only activated when the updated function is called the next time, which is acceptable for online service patching, but meaningless for crash recovery for DL systems. For example, if the training function is terminated to perform updates and then directly re-invoked, actually all of its conducted computation will be lost and thus need to be repeated in its second invocation. This paper proposes \textsc{DaiFu} to fill this gap, and thus enable in-situ recovery for DL systems. It is noteworthy that \textsc{DaiFu} only operates on the user programs, and does not need invasive modifications to the Python interpreter.

\begin{table}[t]
\setlength\tabcolsep{1.5pt}
\caption{A comparison between our work and prior crash recovery methods for DL systems.}
\label{tab:comparison}
\resizebox{1\linewidth}{!}{\begin{tabular}{l|ccccc}
\toprule
\textbf{Methods}& \begin{tabular}[c]{@{}c@{}}\textbf{Support}\\ \textbf{Scenarios}\end{tabular}& \begin{tabular}[c]{@{}c@{}}\textbf{Library}\\ \textbf{Modified}\end{tabular} & \begin{tabular}[c]{@{}c@{}}\textbf{User Code}\\ \textbf{Modified}\end{tabular} & \begin{tabular}[c]{@{}c@{}}\textbf{Retry}\\ \textbf{Granularity}\end{tabular}& \begin{tabular}[c]{@{}c@{}}\textbf{Code}\\\textbf{Update}\end{tabular}\\ \hline
\textbf{Restart} &General & No & No  & Program &Restart\\
 \textbf{Built-in*} & General  & No & Manual  & Epochs& Restart\\
 \textbf{CheckFreq} & General  & Yes & Manual  & Iterations& Restart\\
 \textbf{CheckNRun} & Specific   & Yes & Manual  & Iterations& Restart\\
 \textbf{Gemini} & Specific  & Yes & Manual  & Iterations& Restart\\
 \textbf{JIT-C} & Specific & Yes & Manual  & An Iteration & Restart\\
 \cdashline{1-6}
 \textbf{\begin{tabular}[c]{@{}c@{}}\textsc{DaiFu}\\(Our work)\end{tabular}} & General  & No & Semi-Auto  & \begin{tabular}[c]{@{}c@{}}Lines within\\an Iteration\end{tabular} & \begin{tabular}[c]{@{}c@{}}Dynamic\\Update\end{tabular}\\
\bottomrule
\end{tabular}}
\begin{flushleft}
         * DL libraries usually provide built-in support for checkpointing, e.g., torch.save() in PyTorch~\cite{DBLP:conf/nips/PaszkeGMLBCKLGA19}.
\end{flushleft}
\figin
\end{table}

\begin{figure*}[t]
\centering
\subfigure[The components of \textsc{DaiFu}.]{
\label{outline}
\includegraphics[height=0.188\textwidth]{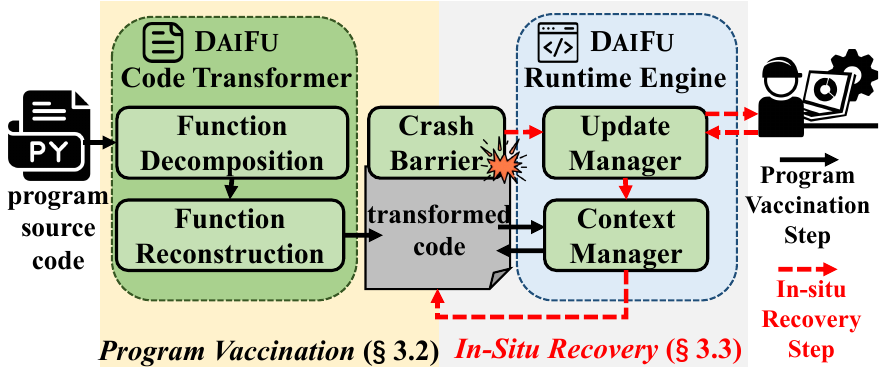}}
\!\hspace{-0.05in}
\subfigure[The working mechanism of \textsc{DaiFu}.]{
\label{outline2}
\includegraphics[height=0.188\textwidth]{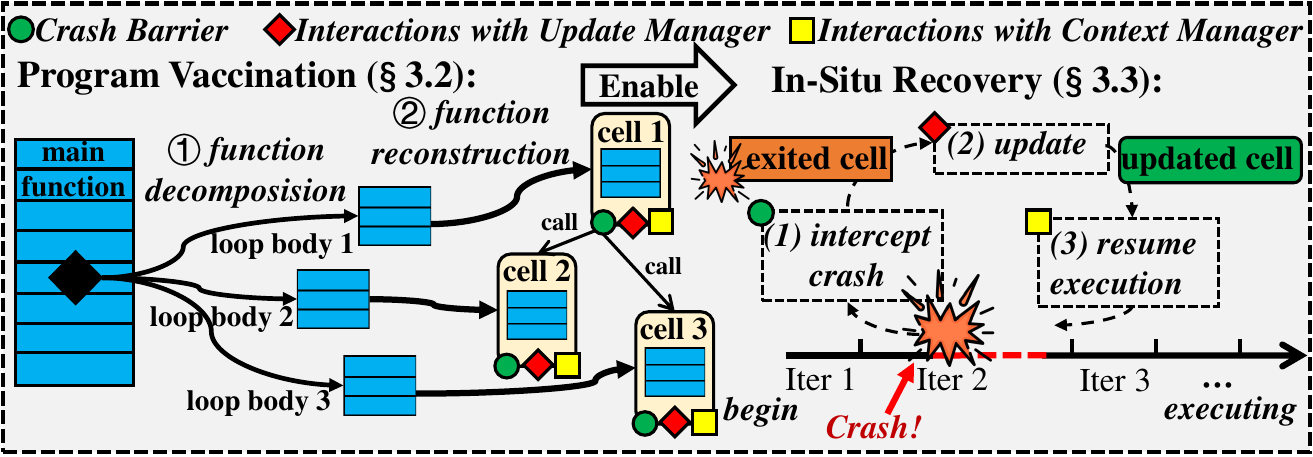}}

\figin
\caption{The overview of \textsc{DaiFu}.}
\label{outline_all}
\end{figure*}

\subsection{Comparison of  \textsc{DaiFu} and Prior Techniques}

Table~\ref{tab:comparison} outlines the key differences between our in-situ recovery framework, namely \textsc{DaiFu}, and the existing crash recovery solutions for DL systems. Specifically, CheckNRun~\cite{DBLP:conf/nsdi/EisenmanMIMKNSA22}, Gemini~\cite{DBLP:conf/sosp/WangJZZFNW23}, and JIT-C~\cite{DBLP:conf/eurosys/GuptaKKVGKRS24} are proprietary tools tailored to specific applications (e.g., embedding table checkpointing for recommender systems or large-scale distributed training) and infrastructures that demand extensive modifications to libraries (e.g., Deepspeed~\cite{deepspeed}, CUDA~\cite{cuda}, NCCL~\cite{nccl}), and they remain closed-source. In contrast, \textsc{DaiFu} is designed for general DL systems. Similarly, CheckFreq~\cite{DBLP:conf/fast/MohanPC21} is coupled with Nvidia DALI~\cite{dali} and requires invasive changes to data loader APIs. Developers must refactor their code to integrate CheckFreq, and its incompatibility with multi-threading libraries, such as failing to work with PyTorch’s “torchrun” for distributed training, adds further challenges. Moreover, prior checkpoint-retry based solutions force developers to manually specify checkpoint targets (e.g., the model, optimizer states, intermediate states, and evaluation results). In contrast, \textsc{DaiFu} automatically transforms the user program once the main function is annotated. Furthermore, by performing in-situ recovery at the code line level, \textsc{DaiFu} significantly reduces the restore time for crash recovery. Finally, unlike previous methods that require complete program restarts for code updates, which incur substantial program initialization overhead, \textsc{DaiFu} can seamlessly apply code updates without program restarts.

\section{\textsc{DaiFu}: Design and Implementation}

\subsection{Overview}

\textsc{DaiFu} is a framework that enables in-situ crash recovery for DL systems. Fig.~\ref{outline_all} presents an overview of \textsc{DaiFu}, which consists of two main processes: program vaccination to prepare for crash recovery, and in-situ recovery if a crash occurs. 
During program vaccination, \textsc{DaiFu} Code Transformer performs function decomposition and function reconstruction, which instrument the program to prepare for in-situ recovery. 
Then, if a program crash occurs during program execution, in-situ recovery will be triggered. \textsc{DaiFu} offers developers interfaces to interact with the vaccinated program and perform dynamic software updates for crash recovery.

\textbf{Usage.} We make the adoption of \textsc{DaiFu} user-friendly.
To integrate \textsc{DaiFu}, developers only need to import it and place a decorator before the entry function (i.e., the main or training function), just as Lines 2 and 4 in Fig.~\ref{example_fault} show. This function is denoted as the to-be-vaccinated function in the following. It is noted that only the entry function needs to be vaccinated since it is long-running, and thus its progress should be carefully preserved. 

\begin{table}[tp]
  \centering
  \caption{The synthesis of unfinished procedure when its beginning occurs at different control flow locations.}
  \figin
  \includegraphics[width=1\linewidth]{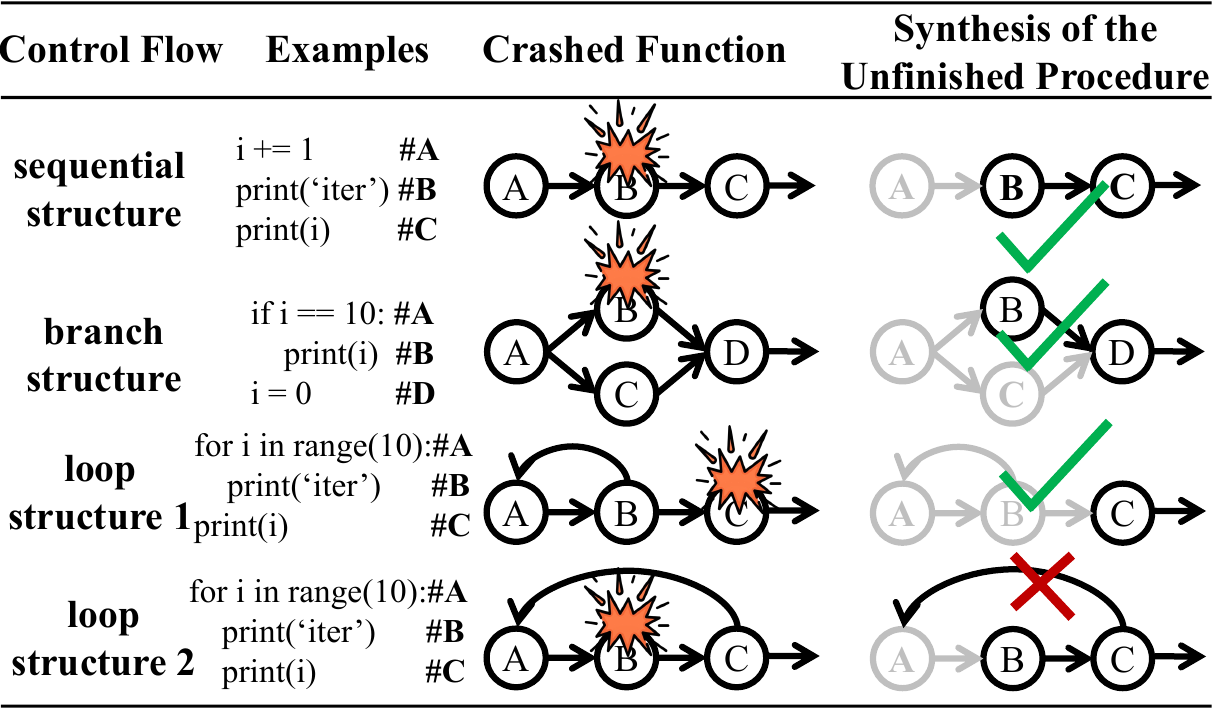}
  \figin
  
\label{control_flow}
\end{table}

\subsection{Program Vaccination}\label{sec:prog_vacc}
As introduced in \S~\ref{sec:dsu}, Python does not allow DSU to active functions. To overcome this challenge, an initial idea is to (1) let the function exit the evaluation stack; (2) catch the thrown exception to avoid a total crash of the program; (3) synthesize a new function that implements the unfinished procedure of the exited function; (4) integrate updates into it; (5) dynamically replace the original function with the new one, and (6) re-execute the new one instead. Here, the steps (2)$\sim$(6) should be prepared in advance with code instrumented into the program by \textsc{DaiFu}. Among these steps, the most difficult step is the step (3), i.e., unfinished procedure synthesis. Table~\ref{control_flow} visualizes how the unfinished procedure synthesis can be achieved in different situations. We can see that, if the unfinished procedure does not begin in a loop (i.e., the sequential structure, branch structure and loop structure~1 in Table~\ref{control_flow}), we can simply skip all the statements executed before the restart location (marked as gray in Table~\ref{control_flow}), and only wrap the remaining statements into the new function for re-execution. However, if the unfinished procedure begins in a loop (i.e., loop structure~2 in Table~\ref{control_flow}), directly skipping all the executed statements is wrong because the control flow contains a back edge pointing to a skipped statement. We consider this situation as an illegal situation, where unfinished procedure synthesis cannot be achieved with a trivial skip. 
To avoid this illegal situation, \textsc{DaiFu} refactors the to-be-vaccinated function. 

\begin{figure*}[t]
  \centering
  \subfigure[An automatic transformation to the training code, whose generated cells are in Fig.~\ref{fig: vacc_c}.]{
  \includegraphics[width=0.75\linewidth]{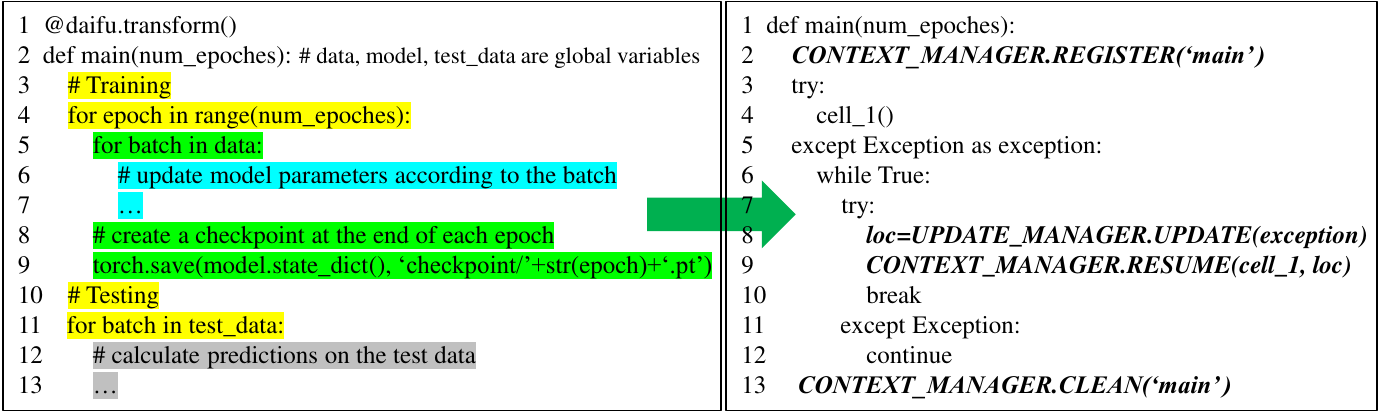}\label{fig: vacc_a}}
  \subfigure[The formulated cell tree.]{
  \includegraphics[width=0.22\linewidth]{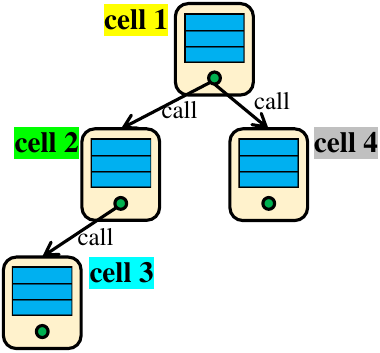}\label{fig: vacc_b}
  }
  \subfigure[The cells transformed from the vaccinated function for execution. The bold italic code are pseudo-code representing how the instrumented context/update managers work.]{
  \includegraphics[width=\linewidth]{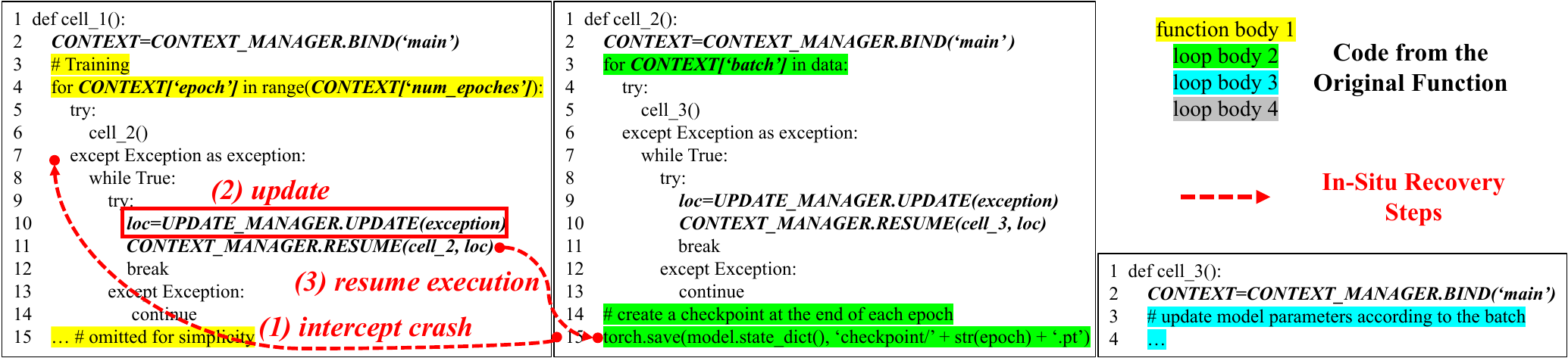}\label{fig: vacc_c}}
  \figin
  \caption{An example of program vaccination, with arrows indicating how the instrumented code enables in-situ recovery.}
  
\label{vacc_example}
\end{figure*}

\textbf{Cell}. Before delving into details of the refactoring process, we first introduce a concept, i.e., cells. Specifically, a cell is: (1) a function does not contain loop structures, or (2) a function contains loop structures, but the body of each loop is a call to another cell wrapped with a crash barrier. Here, a crash barrier is exception handling statements instrumented into the program to prevent exceptions from being thrown out of a loop, and allow the \textsc{DaiFu} Runtime Engine to engage in crash recovery. For example, Fig.~\ref{fig: vacc_c} shows three examples of cells, where \texttt{cell\_3} follows the above condition (1), while \texttt{cell\_1} and \texttt{cell\_2} follow the above condition~(2). The benefit of a cell is that its unfinished procedure will not meet the illegal situation described in loop structure~2 in Table~\ref{control_flow}.

\begin{algorithm}[t]
\caption{Function Decomposition}
\label{func_decompose}
\LinesNumbered
\KwIn {The AST of given function $f\_ast$}
\KwOut {The AST of vaccinated function $v\_ast$; The cell tree $c\_tree$}
Initialize $c\_tree$\\
$v\_ast, c\_tree\gets$\texttt{RecursiveExtract}($f\_ast$, $c\_tree$)\\
\SetKwFunction{FMain}{RecursiveExtract}
\SetKwProg{Fn}{Function}{:}{}
\Fn{\FMain{$node$, $c\_tree$}}{
    \ForEach{child$\in$$node.childrens$}
    {$child, c\_tree \gets$\texttt{RecursiveExtract}($child$, $c\_tree$)\\
    \If {child.type $\in$ \{$'For'$, $'While'$, $'FuncDef'$\}}{
        $cell\gets$$AST\_Node('FuncDef', child.body)$\\
        $call\gets$$AST\_Node('Call', cell)$\\
        $barrier$$\gets$$AST\_Node('Try', call)$\\
        Add crash handling code into $barrier.handler$\\
        $child.body\gets$$barrier$\\
        $c\_tree.insert\_child(cell)$}
    }
}
\Return{node, c\_tree}\\

\end{algorithm}

\begin{algorithm}[t]
\caption{Function Reconstruction}
\label{func_reconstruct}
\LinesNumbered
\KwIn {The AST of vaccinated function $v\_ast$; The cell tree $c\_tree$, the execution namespace $ns$}
\KwOut {The updated execution namespace $ns$}
Instrument namespace registration code into $v\_ast$\\
$v\_dict \gets$ the registered namespace of $v\_ast$\\
\ForEach{cell$\in$$c\_tree$}
{Bind all variables of $cell$ to $v\_dict$\\
Replace break, continue into return a related indicator\\
}
$cell\_nodes \gets$ all AST nodes from all cells from $c\_tree$\\
\ForEach{node$\in$$(v\_ast.traverse()\cup cell\_nodes)$}
{
\If {$node.type$ is $'Call'$ and node.callee $\in$ c\_tree}{
Add code to $node$ to react to the return flag}
}
Initialize and add functions from $v\_ast$ and $c\_tree$ into $ns$\\

\end{algorithm}

\textbf{Function Decomposition}. 
\textsc{DaiFu} refactors the to-be-vaccinated function by decomposing it into a couple of cells. 
Algorithm~\ref{func_decompose} describes this process. Specifically, given the to-be-vaccinated function, \textsc{DaiFu} constructs its cell tree recursively by adding a cell as a child of the current cell when a new loop initialization is identified. The body of the new added cell is the body of the identified loop. Moreover, the call to each cell is wrapped within a crash barrier. The exception handler of the crash barrier includes interactions with the update manager and the context manager, whose mechanism will be illustrated in \S~\ref{sec:fighting}. Fig.~\ref{vacc_example} presents an example of this process. Specifically, the left part of Fig.~\ref{fig: vacc_a} is the original code, and the right part of Fig.~\ref{fig: vacc_a} is its transformed version by \textsc{DaiFu} while the cells decomposed from the function are shown in Fig.~\ref{fig: vacc_c}. With function decomposition, the code of the original function is distributed to different cells, so that the unfinished procedure synthesis can be carried out smoothly.

\textbf{Function Reconstruction}. The semantics of the original function may be broken due to function decomposition. \textsc{DaiFu} avoids this through function reconstruction. Algorithm~\ref{func_reconstruct} describes this process, which includes variable redirection and control flow maintenance. Variable redirection is a process that allows all cells from the same function to share the same namespace. Specifically, the context manager maintains a dictionary of references to variables in the vaccinated function. During function reconstruction, \textsc{DaiFu} Code Transformer instruments each cell to redirect each variable resolution to the dictionary. Besides, since loop bodies are transformed into cell calls, the control flow controlled by ``break'', ``continue'' and ``return'' is broken. To perform control flow maintenance, \textsc{DaiFu} transforms these statements into returning a string indicating the original statement (also with the returned item if the statement is \textit{return}). Then \textsc{DaiFu} adds code after each cell call to react to its returning indicators, so the original control flow can be reconstructed. After function reconstruction, the cells can together perform the original task performed by the vaccinated function, while equipped with the ability to recover in situ from crashes. 

\subsection{In-Situ Recovery}\label{sec:fighting}
We continue to introduce the in-situ recovery process and the support mechanisms for this process. Specifically, when a crash happens, an in-situ recovery consists of 3 steps, 
which are highlighted with red dashed arrows in Fig.~\ref{fig: vacc_c}.

\textbf{(1) Intercept Crash}.
The crash barrier introduced in the former section is responsible for intercepting the crash. The caught exception usually contains some descriptions of the fault that caused the crash, which is helpful for generating an update. 


\textbf{(2) Update}. 
Developers can interact with the program to diagnose the crash first, and then they can perform updates with the interfaces detailed in Table~\ref{tab:primitive}. Specifically, \textsc{DaiFu} supports statement-level retries (with \textit{pass}), code execution to update state such as configuration (with \textit{action}), and in-situ code updates (with \textit{surgery}). 

\begin{table}[t]
\centering
\caption{Interfaces for in-situ recovery by \textsc{DaiFu}.}
\halffigin
\label{tab:primitive}
\resizebox{1\linewidth}{!}{\begin{tabular}{c|p{205pt}}
\toprule
\textbf{Interface} & \textbf{Description}\\\hline
\textit{pass}      &  Taking no input, directly continue the execution of a program exactly from where it fails.           \\
\textit{surgery}   & Taking a piece of code as input, replace the vaccinated function with the input code.            \\
\textit{action}    & Taking a piece of code as input, execute the code instantly within the program context.            \\
\bottomrule
\end{tabular}}
\figin
\end{table}


If the failed function is updated by the \textit{surgery} interface, the update manager will analyze the code changes, find the first modified code location, and use it as a restart location. Otherwise, the update manager will analyze the caught exception, find the code location that throws the exception, and use it as a restart location.

\textbf{(3) Resume Execution}. 
Based on the restart location returned by the update manager, the context manager synthesizes the unfinished procedure of the exited cell as a new function, and then calls it to resume the execution.

\textsc{DaiFu} also provides additional support mechanisms that complement the above steps, which are illustrated below.

\textbf{Support for Distributed Training.} \textsc{DaiFu} supports in-situ recovery for distributed training systems, and the mechanism to support distributed training on multiple servers is the same as supporting distributed training on multiple processes in a server. Specifically, when one or some of the distributed training processes crash, other processes will wait for them due to the synchronization barrier at each iteration. After a developer applies \textsc{DaiFu} to fix one crashing process, the fix procedures are recorded and will be broadcasted to all crashing processes, and the crashing processes will replay the fix procedures. After the crashing processes recover, they can reach the synchronization barrier, and then all processes of the DL system will continue the execution.

\textbf{Support for Root Cause Debugging.} Like all prior DL crash recovery methods~\cite{DBLP:conf/usenix/KimJLLJ22, DBLP:conf/fast/MohanPC21, DBLP:conf/icml/0036LRJ20, DBLP:conf/nsdi/EisenmanMIMKNSA22, DBLP:conf/sosp/WangJZZFNW23, DBLP:conf/eurosys/GuptaKKVGKRS24}, \textsc{DaiFu} still needs developers to perform debugging, but \textsc{DaiFu} makes several efforts to ease this process. Specifically, (1) When a crash is intercepted, \textsc{DaiFu} will print a stack trace annotated with variable values to help developers gain the first view of the crash. (2) We implement \textsc{DaiFu} Repair Manager to inherit from PDB, so developers can interact with the program using PDB commands to understand the crash circumstance. (3) We also implement an automated debugging interface in \textsc{DaiFu}, which can be used to send the variable-annotated stack trace and the program code to a large language model (e.g., OpenAI's GPT), and prompt the model to explain the crash, analyze the root cause, and recommend fixes. The output analysis can help developers better understand a crash.

\section{Benchmark Construction}\label{sec:what_happen}

A benchmark of executable crash scenarios in DL systems is needed to evaluate \textsc{DaiFu}, whose desired properties include: 
\begin{itemize}[leftmargin=*]
\item \textbf{Reproducibility.} Reproducing crashes of DL systems is non-trivial. According to a study~\cite{DBLP:journals/corr/abs-2206-12311}, only about 5.3\% of all reviewed DL program faults in GitHub can be reproduced. We accumulate these valuable cases from existing benchmarks~\cite{DBLP:journals/corr/abs-2206-12311, DBLP:conf/icse/LiangLSSFD22}.
\item \textbf{Good Coverage on Diverse Crash Scenarios.} 
Covering more crash scenarios discussed in prior studies~\cite{DBLP:conf/sc/Hu0Y0021, DBLP:conf/usenix/JeonVPQXY19} into the benchmark can facilitate a more comprehensive evaluation.
\item \textbf{Inclusion of Long-Running DL Systems.} The benchmark needs to include some real-world DL systems that operate on large-scale data and execute for a long time.
\end{itemize}

\textbf{Method.} We first search for reproducible crashing faults of DL systems from prior empirical studies~\cite{DBLP:conf/issta/ZhangCCXZ18, DBLP:conf/icse/IslamPNR20, DBLP:conf/icse/HumbatovaJBR0T20, DBLP:journals/corr/abs-2206-12311, DBLP:conf/icse/LiangLSSFD22} and technique studies~\cite{DBLP:journals/tosem/NikanjamBMK22, DBLP:conf/icse/WardatLR21, DBLP:conf/icse/ZhangZMS21, DBLP:conf/icse/CaoLC0TWC22}. For each study, we first identify whether the associated artifact provides scripts to reproduce the faults reported in the paper. Then, for each reproducible fault, we confirm whether it will cause a crash. We also validate that these cases will escape the sanity checks~\cite{pyright} by VS Code to omit trivial cases. Finally, to cover more crash scenarios, we inject crashing faults to some representative DL systems to enrich the benchmark. 

\textbf{Reproducing Real-World Crashing Faults}. Based on the exploration, some cases are excluded from our benchmark due to the lack of reproduction information~\cite{DBLP:conf/issta/ZhangCCXZ18, DBLP:conf/icse/IslamPNR20, DBLP:conf/icse/HumbatovaJBR0T20} or a mismatch of subjects (i.e., the technique studies~\cite{DBLP:conf/icse/ZhangZMS21, DBLP:conf/icse/WardatLR21, DBLP:conf/icse/CaoLC0TWC22, DBLP:journals/tosem/NikanjamBMK22} do not provide faults leading to a crash). 
Finally, the successfully-reproduced crash scenarios come mainly from defects4ML~\cite{DBLP:journals/corr/abs-2206-12311} and gDefectDL~\cite{DBLP:conf/icse/LiangLSSFD22}. From these benchmarks, we select real-world faults that will cause a crash and remove duplicate cases. 
Finally, we reproduced a total of 18 crash cases in real-world DL systems.



\begin{table}[t]
\centering
\caption{Some representative DL systems in our benchmark.}
\halffigin
\label{long_running_subject}
\resizebox{1\linewidth}{!}{\begin{tabular}{llll}
\toprule
\textbf{Domain}       & \textbf{Dataset}   & \textbf{Model} \\ \midrule
\multirow{3}{*}{Image}    & \multirow{3}{*}{ImageNet~\cite{DBLP:journals/ijcv/RussakovskyDSKS15} (146GB)} & ResNet50~\cite{DBLP:conf/cvpr/HeZRS16} (195.34MB)         \\
                          &                                   & ViT-L/16~\cite{DBLP:conf/iclr/DosovitskiyB0WZ21} (2.27GB)           \\
                          &                                   & Swin-B~\cite{DBLP:conf/iccv/LiuL00W0LG21} (670.28MB)           \\
\multirow{2}{*}{Language}                  & OpenWebText~\cite{Gokaslan2019OpenWeb} (54 GB)               & GPT2~\cite{radford2019language} (1.44GB)                 \\
&       Alpaca~\cite{alpaca} (133MB)         & LLaMA-7B~\cite{DBLP:journals/corr/abs-2302-13971} (25.10GB)                 \\
\bottomrule
\end{tabular}}
\end{table}

\textbf{Including More Crash Scenarios with Fault Injection.} 
We further enrich our benchmark via fault injection. We first select some representative DL systems, including three demo programs~\cite{pytorch_examples} from Computer Vision (CV), Natural Language Processing (NLP) and Reinforcement Learning (RL), and five long-running DL systems including training on ImageNet~\cite{DBLP:journals/ijcv/RussakovskyDSKS15} for ResNet50~\cite{DBLP:conf/cvpr/HeZRS16}, ViT-L/16~\cite{DBLP:conf/iclr/DosovitskiyB0WZ21} and Swin-B~\cite{DBLP:conf/iccv/LiuL00W0LG21}, and training on OpenWebText~\cite{Gokaslan2019OpenWeb} for GPT2~\cite{radford2019language}, and finetuning on Alpaca~\cite{alpaca} for LLaMA-7B~\cite{DBLP:journals/corr/abs-2302-13971}. 
The short-running base DL systems are used to efficiently evaluate the usefulness of \textsc{DaiFu} in different crash scenarios, while the long-running DL systems are used for controlled experiments to understand the practical effect and benefit of different crash recovery techniques. 
Specifically, we inject faults into the source code, runtime environment, or loaded data to cover more crash scenarios summarized in recent studies~\cite{DBLP:conf/icse/ZhangXZLLY20, DBLP:conf/usenix/JeonVPQXY19}. For example, to simulate Runtime Error, we select some device-related APIs (e.g., \texttt{to} for GPU-CPU operation, \texttt{save} for model persistence, and \texttt{all\_reduce} or \texttt{broadcast\_coalesced} for distributed training), and inject code to control them to throw an exception. 
By doing so, the symptom is similar to real-world crash cases caused by transient runtime errors (i.e., the system crashes after interacting with some devices).

\textbf{Summary.} Finally, the benchmark contains DL systems with crash scenarios caused by various root causes. The benchmark contains a total of 32 crash cases from 7 crash scenarios. The description of each crash scenario is presented in Table~\ref{tab:crash_scenrio}.

\section{Evaluation}\label{sec:evaluation}
Our evaluation aims to answer the following questions: 
\begin{itemize}[leftmargin=*]
    \item \textbf{RQ1 (\S~\ref{sec:rq1}):} How much can \textsc{DaiFu} help speed up crash recovery for DL systems? 
    \item \textbf{RQ2 (\S~\ref{sec:rq2}):} What runtime overhead does \textsc{DaiFu} impose to its vaccinated DL systems?  
    \item \textbf{RQ3 (\S~\ref{sec:rq3}):} How many crash scenarios are supported by \textsc{DaiFu}?
    \item \textbf{RQ4 (\S~\ref{sec:rq4}):} Do DL systems vaccinated by \textsc{DaiFu} still function correctly after in-situ recovery? 
\end{itemize}




\begin{table}[t]
\centering
\caption{Crash scenarios included in our benchmark.}
\halffigin
\label{tab:crash_scenrio}
\resizebox{1\linewidth}{!}{\begin{tabular}{l|l|c}
\toprule
\textbf{Scenario} & \textbf{Description} & \#\\\hline
\textbf{\textit{API Misuse}} &  DL library functions are used incorrectly. &  12        \\ 
\textbf{\textit{Tensor Mismatch}} & Tensor shapes or data types are incompatible. &  6     \\ 
\textbf{\textit{Resource Bug}} &  Excessive GPU memory allocation is required. &  3   \\ 
\textbf{\textit{GPU Contention}} &  Hosted GPUs are occupied by other programs. &  3      \\ 
\textbf{\textit{Path Problem}} &  Specified file or directory paths are incorrect. & 3         \\ 
\textbf{\textit{Exceptional Data}} & The input data are unexpected or anomalous. &  2    \\ 
\textbf{\textit{Runtime Error}} & Temporary issues like network instability occur. & 3         \\ 
\bottomrule
\end{tabular}}
\figin
\end{table}

\begin{table*}[tb] \centering
    \renewcommand{\arraystretch}{0.9}
    \caption{Comparisons of different recovery methods on the averaged restore time for crash recovery, and runtime overhead.}
    \figin
    \label{tab:recovery_and_oh}
    \resizebox{0.75\linewidth}{!}{
        \begin{tabular}{l|*{2}{c}|*{2}{c}|*{2}{c}|*{2}{c}|*{2}{c}}
        \toprule
   \multirow{2}{*}{\textbf{Methods}}  & \multicolumn{2}{c}{ResNet50~\cite{DBLP:conf/cvpr/HeZRS16}} 
                               & \multicolumn{2}{c}{ViT-L/16~\cite{DBLP:conf/iclr/DosovitskiyB0WZ21}} 
                               & \multicolumn{2}{c}{Swin-B~\cite{DBLP:conf/iccv/LiuL00W0LG21}} 
                               & \multicolumn{2}{c}{GPT2~\cite{radford2019language}}  
                               & \multicolumn{2}{c}{LLaMA-7B~\cite{DBLP:journals/corr/abs-2302-13971}}  \\
                               & \textbf{RT*} & \textbf{OH*} 
                               & \textbf{RT*} & \textbf{OH*} 
                               & \textbf{RT*} & \textbf{OH*} 
                               & \textbf{RT*} & \textbf{OH*} 
                               & \textbf{RT*} & \textbf{OH*}\\
        \midrule
 \textbf{Restart} & 2393.65s & 0\%& 16370.14s & 0\%& 5618.54s & 0\%&12026.00s & 0\%& 10018.27s & 0\%\\
   \textbf{Original}  &771.32s&0\%&5414.86s&0\%&1860.14s&0\%&3921.67s&0\%&2532.97s&0\%\\
    \textbf{CheckFreq}  &440.12s&7.64\%&2692.10s&3.67\%&942.18s&4.53\%&\multicolumn{2}{c|}{ RuntimeError**}&363.00s&2.75\%\\
    \textbf{\textsc{DaiFu}}  &0.28s&0.40\%&2.16s&0.03\%&0.74s&$\approx$0\%&3.97s&$\approx$0\%&0.30s&$\approx$0\%\\
        \bottomrule
    \end{tabular}}
    \begin{flushleft}
         * RT denotes the Restore Time for crash recovery, OH denotes the extra Runtime Overhead during normal execution.\\
        ** This runtime error is due to the incompatibility issue between CheckFreq and the employed torchrun in GPT2. 
    \end{flushleft}
    \figin
\end{table*}

\subsection{Setup}
\textbf{Benchmark Setup.} We carry out experiments using the benchmark introduced in \S~\ref{sec:what_happen}. Specifically, in RQ1 (\S~\ref{sec:rq1}) and RQ2 (\S~\ref{sec:rq2}), we use \textsc{DaiFu} to train ResNet50~\cite{DBLP:conf/cvpr/HeZRS16}, ViT-L/16~\cite{DBLP:conf/iclr/DosovitskiyB0WZ21}, Swin-B~\cite{DBLP:conf/iccv/LiuL00W0LG21} and GPT2~\cite{radford2019language} with distributed data parallelism on 4 V100 GPUs, and fine-tune LLaMA-7B~\cite{DBLP:journals/corr/abs-2302-13971} on an A100 GPU. The number of training epoches of ResNet50, ViT-L/16 and Swin-B are set as 3, and the number of training iterations of GPT2 and LLaMA-7B are set as 6000 and 62500 respectively. For other configurations, we use the default values provided in their code.  For RQ1 and RQ2, we implement a controller to trigger injected faults routinely during the execution of these DL systems. Through these fault injections, we can understand the effectiveness of different methods to recover crashes occurring at different time. Then in RQ3 (\S~\ref{sec:rq3}) and RQ4 (\S~\ref{sec:rq4}), we will further evaluate and discuss the usefulness and correctness of \textsc{DaiFu} for different crash scenarios.

\textbf{Method.} 
For each case, we find the main, training and evaluation functions, and then decorate them for vaccination. If a DL system does not place training and evaluation code in a function but places them in the global namespace, we wrap the code into a function for vaccination, and then invoke the vaccinated function instead. During the experiments, we monitor the program execution and the results to evaluate \textsc{DaiFu}. The recovery history of each case is provided in our artifact~\cite{daifu_artifact} to facilitate reproduction.


\textbf{Experiment Environment.} We run our experiments in two servers, including one server with two 2.8GHz 16-core Intel Xeon Gold 6242 CPUs, 4 NVIDIA V100 GPUs, and 128GB of RAM, and the other server with four 2.1GHz 12-core Intel Xeon Silver 4116 CPU, 1 NVIDIA A100 GPU, and 128GB of RAM. We use Python 3.6 to 3.11 for experiments, depending on the specific Python version required by each case.

\textbf{Baselines.} 
We adopt the following methods for comparison.

\begin{itemize}[leftmargin=*]
\item \textbf{Restart.} The crashed DL system is repaired and then restarted from scratch to recover from a crash. This direct approach is a widely-applied practice in prior work~\cite{DBLP:conf/icse/CaoLC0TWC22, DBLP:conf/icse/WardatLR21, DBLP:conf/chi/SchoopHH21, DBLP:conf/icse/ZhangZMS21, DBLP:conf/icse/WardatCLR22, DBLP:conf/icse/LiZRXX23}.
\item \textbf{Original.} The DL systems in our benchmark are open-source projects, which also contain manually-implemented checkpoint-retry. The corresponding code is written by their developers, we use this original checkpoint-retry code as one of the baselines.
\item \textbf{CheckFreq \cite{DBLP:conf/fast/MohanPC21}.} CheckFreq is a state-of-the-art DL crash recovery method based on asynchronous checkpointing. We adopt 
its open-source implementation with its default configuration, and modify the code of each benchmark program to use it. 

\end{itemize}

There are also other checkpoint-retry based methods~\cite{DBLP:conf/nsdi/EisenmanMIMKNSA22, DBLP:conf/sosp/WangJZZFNW23, DBLP:conf/eurosys/GuptaKKVGKRS24}, but they are closed-source, and only optimize checkpoint-retry in specific situations (e.g., CheckNRun~\cite{DBLP:conf/nsdi/EisenmanMIMKNSA22} for embedding tables in recommender systems and Gemini~\cite{DBLP:conf/sosp/WangJZZFNW23} for checkpoint placement in distributed training), yet the basic idea is similar. Only CheckFreq is publicly available, so we select it as a baseline here.

\subsection{RQ1: Crash Recovery Speedup}\label{sec:rq1}

We compare different DL crash recovery methods based on how much time is spent on restoration for crash recovery with these methods. Since the occurrence time of crashes also affects the restore time for crash recovery, to ensure a fair comparison, we consistently inject runtime errors at identical training iterations for experiments of different recovery methods. The injections are uniformly distributed throughout the entire training process. 
We write scripts to automatically repeat the recovery process for each recovery method. The experiments for each recovery method are repeated 3 times. 
The time elapsed between the occurrence of the crash and the completion of recovery is calculated as the restore time. Specifically, we consider a recovery to be completed when the program arrives at the end of its original crashing iteration, so we record the recovery completion time as this moment.
\begin{figure}[t]
  \centering
  \includegraphics[width=1\linewidth]{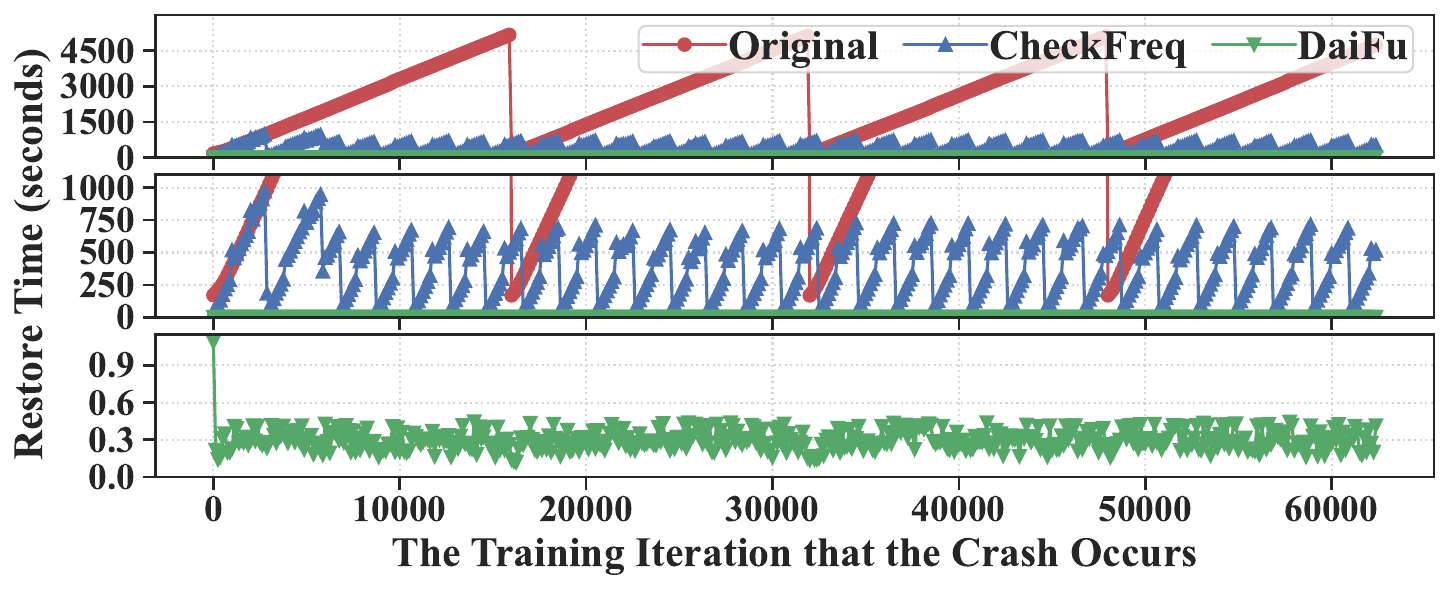}
  \figin
  \caption{The restore time for crashes occurring at different time when finetuning LLaMA with different DL crash recovery methods. We display the comparison under different scales to provide a clear comparison.}
  \figin
\label{example_llama}
\end{figure}

The columns ``RT*'' in Table~\ref{tab:recovery_and_oh} display the average restore time achieved by different methods. We can see that \textsc{DaiFu} can consistently achieve restoration for crash recovery in a few seconds (i.e., 0.28s-3.97s), achieving a speedup (i.e., $\frac{RT\ of\ Baseline}{RT\ of\ DaiFu}$) of over 1327$\times$ compared to other baselines. The superior recovery speedup by \textsc{DaiFu} is originated from its in-situ recovery mechanism, which eliminates the need to repeat the program warm-up (e.g., data loading) and the lengthy training process between the crash and the latest checkpoint or the program starting point.

To better understand the difference between \textsc{DaiFu} and prior solutions, we visualize their achieved restore time for crashes occurring at different training iterations. In Fig.~\ref{example_llama}, we can see that the scale difference between the restore time achieved by \textsc{DaiFu} and these baseline solutions is significant. Specifically, \textsc{DaiFu} almost continuously achieves restoration in 1 second for LLaMA-7B training, but the restore time achieved by other crash recovery methods can be up to more than 4500s. In a nutshell, if a crash can be intercepted and recovered, in-situ recovery can stably provide a faster recovery compared to other crash recovery techniques.

In summary, \textsc{DaiFu} can provide quick crash recovery for DL systems. With this ability, not only can computing time and resources be saved, but developers can also get timely debugging feedback after taking some measures to recover a system from a crash.

\subsection{RQ2: Overhead}\label{sec:rq2}

We continue to evaluate the runtime overhead incurred by \textsc{DaiFu} and CheckFreq during normal executions. Specifically, we record the execution time $T_e$ of the recovery-enhanced version (i.e., armed with \textsc{DaiFu} and CheckFreq) and the execution time $T_o$ of the original version. Then we calculate their relative difference as the \textit{runtime overhead}, i.e., $\frac{T_e-T_o}{T_o}$. The experiments are repeated 3 times for each recovery method, and the average runtime overhead is recorded. During experiments, we observe that the overhead of \textsc{DaiFu} is close to 0\% and therefore cannot be precisely represented sometimes, so we mark these small values as $\approx$0\% during recording. 


\begin{table}[t]
\centering
\caption{Breakdown of the \textsc{DaiFu} runtime overhead.}
\figin
\label{tab:overhead_breakdown}
\begin{tabular}{llc}
\toprule
\multicolumn{2}{l}{\textbf{Process}}                              & \textbf{Extra Time} \\ \midrule
\multicolumn{2}{l}{Print Statement (As a Baseline)}                         & 2.04e-06s          \\
\multicolumn{2}{l}{\textbf{Crash Barrier}}         & 6.54e-08s           \\
\multirow{2}{*}{\textbf{Variable Redirection}} & \textbf{-Bind}   & 8.69e-08s          \\
                                                & \textbf{-Resolution} & 1.65e-09s           \\
\bottomrule
\end{tabular}
\figin
\end{table}

As shown in the columns ``OH*'' in Table \ref{tab:recovery_and_oh}, the runtime overhead incurred by \textsc{DaiFu} is consistently lower than the overhead incurred by CheckFreq, with a drop rate of over 98\%. The overhead of \textsc{DaiFu} is low and can even be neglected (i.e., $\approx$0\%-0.40\%). This is because compared with adding IO pipeline operations to enable frequent checkpoints, the added operations of \textsc{DaiFu} to the vaccinated programs during normal executions are much more lightweight.

Specifically, the overhead of \textsc{DaiFu} during normal program execution is primarily incurred by the crash barriers and the variable redirection operations (including variable binding and resolution). To understand how these operations affect the execution time, we also break down the overhead of \textsc{DaiFu} by repeating experiments on some microbenchmarks and calculating the extra execution time of each of these operations. Their execution time and one baseline operation for comparison are presented in Table~\ref{tab:overhead_breakdown}. We can see that the added operations by \textsc{DaiFu} generally incur a very low overhead, which is even much lower than that of a print statement. Even though they need to be executed thousands of times during program execution, the accumulated overhead can still be neglected, which is verified in the end-to-end overhead result in Table~\ref{tab:recovery_and_oh}. 

In summary, using \textsc{DaiFu} to enable in-situ recovery incurs a low overhead to the normal execution of the DL system.



\subsection{RQ3: Applicability to Different Scenarios}\label{sec:rq3}
We use \textsc{DaiFu} in different crash scenarios included in the benchmark to examine the applicability of \textsc{DaiFu} in different situations. Detailed recovery actions are generated based on the patch associated with each collected crash case. To evaluate \textsc{DaiFu}'s applicability to different crash cases, we place a log statement at the end of each vaccinated program. Then, we record whether each vaccinated program can normally reach its end after its crashes are recovered in situ with \textsc{DaiFu}.
\subsubsection{Overall Result}
As shown in Table~\ref{tab:overall_result}, \textsc{DaiFu} is applicable to 7 distinct crash scenarios (including 31/32 of the crash cases in the benchmark). Specifically, \textsc{DaiFu} is particularly useful when developers need to frequently update their programs (e.g., add more training data, add more evaluations, or implement new training strategies) in the early stage of program development. In this stage, programming errors are frequent and can seriously affect the productivity of developers, while with \textsc{DaiFu}, the development process can be accelerated. In general, these results demonstrate that the in-situ recovery by \textsc{DaiFu} is applicable in many scenarios.

\begin{table}[t]
\caption{The success rate of recovering different crash cases in situ with \textsc{DaiFu} and its variants. More details of each crash case and the recovery actions are provided in our artifact~\cite{daifu_artifact}.}
\halffigin
\label{tab:overall_result}
\resizebox{\linewidth}{!}{\begin{tabular}{l|c|c|c}
\toprule
\textbf{Scenarios} & \textbf{\textsc{DaiFu} w/o DSU} & \textbf{\textsc{DaiFu} w/o FD} & \textbf{\textsc{DaiFu}} \\ \hline
\textbf{\textit{API Misuse}} & 0/12 & 4/12 & 11/12 \\ 
\textbf{\textit{Tensor Mismatch}} & 0/6 & 3/6 & 6/6 \\
\textbf{\textit{Resource Bug}} & 0/3 & 3/3 & 3/3 \\
\textbf{\textit{GPU Contention}} & 0/3 & 3/3 & 3/3 \\
\textbf{\textit{Path Problem}} & 0/3 & 1/3 & 3/3 \\
\textbf{\textit{Exceptional Data}} & 0/2 & 0/2 & 2/2 \\
\textbf{\textit{Runtime Error}} & 3/3 & 3/3 & 3/3 \\ \hline
\textbf{Summary} & \textbf{3/32 (9.38\%)} & \textbf{17/32 (53.13\%)} & \textbf{31/32 (96.88\%)}\\
\bottomrule
\end{tabular}}
\halffigin
\end{table}

\subsubsection{Ablation Study}
We further investigate whether the inclusion of DSU and Function Decomposition (FD) helps \textsc{DaiFu} support more crash scenarios. Table~\ref{tab:overall_result} displays the results of different variants of \textsc{DaiFu}.  Specifically, \textsc{DaiFu} w/o DSU can only recover crashes caused by transient runtime errors, since only these crashes can be recovered with raw retries without software updates. Incorporating DSU can help recover crashes whose recovery needs software updates, but without function decomposition, the allowed updates are restricted to short-running functions. With function decomposition, \textsc{DaiFu} allows dynamic software updates to the long-running entry function, thus supporting crash recovery in more scenarios.

\subsubsection{Case Study} 
To better understand how \textsc{DaiFu} can be used to perform in-situ crash recovery for DL systems, we analyze some cases in detail. More details of each case are in our artifact~\cite{daifu_artifact}. 

\textbf{Success Case I: API Misuse~\cite{chingyaoc/ggnn.pytorch/9c58ca6}}. This is a crash caused by a misunderstanding of the return value of an API. As shown in Fig.~\ref{casei}, the buggy program in the crashing DL system treats a calculated loss value (0-dim tensor) as a 1-dim tensor and then accesses it with a non-existent index, inducing an error message ``IndexError: invalid index of a 0-dim tensor. Fortunately, the error message thrown by PyTorch directly points out how to fix such a crash. One can fix it in situ with \textsc{DaiFu} using the \textit{surgery} interface, by providing a correct implementation of the suffering test function. Then \textsc{DaiFu} can update this correct version to the program and continue to execute it.


\textbf{Success Cases II: GPU Contention~\cite{daifu_artifact}}. These are crashes caused by GPU contention. In each case, the crashing DL system is hosted on a shared GPU, and during its execution, the GPU is gradually occupied by other programs. As a result, when the DL system needs to apply for new GPU memory but no GPU memory is left, it will throw a runtime error ``RuntimeError: CUDA out of memory''. Fortunately, there may sometimes be other GPUs as alternatives, so we can shift all the tensors in the DL system to another GPU to recover this crash. Specifically, for DL systems implemented with PyTorch, one can use the \texttt{torch.Tensor.to} API to set a new device for each tensor-type variable. This can be executed in situ via the \textit{action} interface, and then \textsc{DaiFu} can continue the program on a device with sufficient resources.




\textbf{Failed Case: API Misuse~\cite{tensorflow/models/9d96e9f}}. This crash is caused by a leaky reuse of the variable scope (an API used for variable sharing in TensorFlow), with an error message ``Value Error: Variable xxx does not exist''. In the fault-free version of this case, the developer corrects the problem by explicitly initializing a new variable scope. This fix solution cannot be directly transformed to an in-situ fix solution since the part to be re-executed (between the variable scope initialization and the crashing location) is not idempotent, and thus cannot update the wrong program state. As a result, the negative effect of the fault cannot be eliminated by \textsc{DaiFu} with a simple re-execution of the revised code. We will explore a fault effect elimination solution for such cases in the future.

\begin{figure}[t]
\begin{lstlisting}[language=diff]
     output = net(init_input, annotation, adj_matrix)
-    test_loss += criterion(output, target).data[0]
+    test_loss += criterion(output, target).data.item()
     correct += pred.eq(target.data.view_as(pred)).cpu().sum()
\end{lstlisting}
\halffigin
\caption{The software updates in the Success Case I.}
\label{casei}
\figin
\end{figure}

\subsection{RQ4: Correctness Tests}\label{sec:rq4}
To test whether \textsc{DaiFu} affects the program correctness, we compare the outcome (e.g., the classification accuracy) of a DL system recovered by \textsc{DaiFu} with that recovered by Restart-from-Scratch. If the difference between two outcomes is not statistically significant, we consider that \textsc{DaiFu} passes the test. Specifically, for each crash case associated with an evaluation measurement, we repeatedly exercise \textsc{DaiFu} for 10 times. Then, following prior work~\cite{DBLP:conf/issta/HumbatovaJT21, DBLP:conf/icst/JahangirovaT20}, we calculate the statistical significance with generalized linear model~\cite{nelder1972generalized}.

In our benchmark, \textsc{DaiFu} passes all the correctness tests. This demonstrates that, the in-situ recovery aided by \textsc{DaiFu} does not harm the original semantics of the vaccinated DL system. 



\section{Discussion}

\subsection{Preparation Effort}\label{sec:effort}

{The preparation effort} refers to the modifications needed to enable crash recovery. Checkpoint-retry based methods require developers to decide which elements to save (e.g., model state, optimizer state, and temporary variables) and how to resume from these checkpoints. More frequent checkpointing further increases the burden, as these methods are often intertwined with underlying deep learning or driver libraries. Consequently, developers must adjust the runtime environment and rewrite portions of their code to accommodate specific API versions (for example, a re-implemented DALI dataloader in CheckFreq~\cite{DBLP:conf/fast/MohanPC21}). In contrast, \textsc{DaiFu} adopts an independent approach that is not built upon any DL libraries. Our experiments demonstrate its successful integration with systems implemented in PyTorch, TensorFlow, and PyTorch Lightning. Moreover, the automated code transformation reduces modifications to only a few clean lines, which is a significant advantage.

\subsection{Compatibility with Third-Party Libraries}

Currently, \textsc{DaiFu} is compatible with many third-party libraries, since in most situations, \textsc{DaiFu} works by transforming the user code and does not interfere with the libraries used. A possible incompatibility is that some libraries provide APIs to encapsulate all training progress, so if such APIs are used in a DL system, \textsc{DaiFu} should be used to vaccinate these APIs to prepare for in-situ recovery. Another possible incompatibility is that, when models are compiled using techniques such as PyTorch JIT~\cite{pytorch_jit} or TensorFlow static graph mode, some of the control flow is transferred to lower-level non-Python runtime, and consequently, the in-situ recovery ability provided by \textsc{DaiFu} is restricted. In these cases, developers may need to disable such compilation features to allow \textsc{DaiFu} to function as intended, or restrict dynamic software updates to other parts of Python code. In the future, we will continue to develop \textsc{DaiFu} to make it compatible with more other libraries.

\subsection{Limitation}\label{sec:limitation}


The limitation of \textsc{DaiFu} is that some crashes cannot be recovered in situ. Specifically, (1) \textsc{DaiFu} cannot recover crashes that cannot be intercepted. For example, a DL system terminated with a ``kill -9'' by other programs cannot be recovered in situ since the process disappears immediately when the crash occurs. 
(2) \textsc{DaiFu} cannot recover crashes that will corrupt the program context and cause an irreversible effect. Typically, these crashes are caused by some non-idempotent harmful logic faults. An example is a fault that clears the model state or deletes the training data. 
In this situation, even after the fault has been removed, the negative effect of the original faulty execution cannot be eliminated via an in-situ retry.  However, even though \textsc{DaiFu} is not available for some crashes, it is still effective in many scenarios as presented in Table~\ref{tab:overall_result}. The main objective of \textsc{DaiFu} is not to replace prior crash recovery paradigms. In fact, \textsc{DaiFu} is synergistic with prior approaches. Benefited from its fast recovery ability for many crash scenarios, with low overhead and small preparation effort, \textsc{DaiFu} is useful enough, and can be applied together with prior approaches to achieve better crash recovery.

\subsection{Threats to Validity}

\textbf{The threat to external validity} lies mainly in the generalizability of our findings to a broader range of DL applications and crash scenarios. To mitigate this threat, we have revisited existing DL fault benchmarks and tried to reproduce as many crash scenarios as possible. We also inject faults into programs in real-world DL systems, to induce crashes with root causes which are not included in prior benchmarks. Through this process, we have included common DL crash scenarios \cite{DBLP:conf/icse/ZhangXZLLY20} into the benchmark. In the future, we will validate \textsc{DaiFu} in more scenarios from large-scale production environments.

\textbf{The threat to internal validity} lies mainly in the completeness of our implementation of \textsc{DaiFu}. To reduce this threat, we have performed multiple rounds of debugging, and carefully tested the code (e.g., the correctness tests in \S~\ref{sec:rq4}) to ensure its quality.

\section{Related Work}

\textbf{Empirical Studies of Real-World Faults in DL Systems.} Plenty of prior studies~\cite{DBLP:conf/icse/ZhangXZLLY20, DBLP:conf/usenix/JeonVPQXY19, DBLP:conf/issta/ZhangCCXZ18, DBLP:conf/icse/IslamPNR20, DBLP:conf/icse/HumbatovaJBR0T20} conduct empirical analysis of faults in DL systems. However, most of these studies primarily focus on classifying faults and do not consider the crucial aspect of their reproduction. In this paper, we aim to fill this gap by constructing a reproducible benchmark of crashing faults in DL systems based on insights from previous studies. Our benchmark encompasses diverse types of crash scenarios and serves as a valuable resource for research on crash recovery for DL systems. 


\textbf{Crash Recovery for DL Systems.} 
To reduce the restore time for crash recovery, prior work~\cite{DBLP:conf/usenix/KimJLLJ22, DBLP:conf/fast/MohanPC21, DBLP:conf/icml/0036LRJ20, DBLP:conf/nsdi/EisenmanMIMKNSA22, DBLP:conf/sosp/WangJZZFNW23, DBLP:conf/eurosys/GuptaKKVGKRS24} optimizes the \textit{Checkpoint-Retry} mechanism to achieve crash recovery for DL systems. For example, CheckFreq~\cite{DBLP:conf/fast/MohanPC21} proposes a PyTorch checkpointing framework that can pipeline checkpointing with computation, and provide algorithmic adjustment of the checkpointing frequency. Despite the benefits of the \textit{Checkpoint-Retry} mechanism in reducing restore time, frequent checkpointing inevitably results in high runtime overhead and increased preparation effort. Different from prior work, this paper introduces \textit{In-Situ Recovery} rather than optimizing existing \textit{Checkpoint-Retry} mechanisms. Benefited from in-situ recovery, \textsc{DaiFu} achieves fast recovery in many crash scenarios, while incurring low overhead and preparation effort. 

\textbf{DSU Techniques.} Numerous DSU techniques have been developed for online services implemented with Python~\cite{DBLP:conf/usenix/HuangXZ021, DBLP:conf/apsec/TangZ18}, C/C++ \cite{DBLP:conf/pldi/NeamtiuHSO06, DBLP:conf/icse/ChenYCZY07, DBLP:conf/usenix/HjalmtyssonG98, DBLP:conf/osdi/RommelDFKBMSL20} or Java \cite{DBLP:conf/eurosys/NicoaraAR08, DBLP:conf/pldi/SubramanianHM09, DBLP:conf/oopsla/PinaVH14}, but they are not applicable to crash recovery for DL systems. 
First, these techniques cannot intercept crashes.
Second, unlike online services, where fault-prone logic is often implemented in functions that process user requests and can exit and then be updated multiple times as the server program is alive, the fault-prone logic in a DL system may be implemented in a function spanning the program life-cycle, and is difficult to update since it cannot exit during execution. As a result, previous DSU techniques developed for Python cannot be applied to patch these functions in DL systems. \textsc{DaiFu}, on the other hand, fills this gap by proposing program vaccination to enable partial exits and dynamic updates of the active entry function of a DL system.

\section{Conclusion}
This paper presents \textsc{DaiFu}, the first technique that enables in-situ crash recovery for DL systems. 
\textsc{DaiFu} prepares in-situ crash recovery for a DL system by decomposing and reconstructing its long-running entry function in advance. The preparation allows the DL system to be dynamically and instantly updated for crash recovery without program termination.
Our promising results demonstrate that \textsc{DaiFu} can 
accelerate the crash recovery practice. Moreover, the overhead imposed by \textsc{DaiFu} is low (under 0.40\%). We believe \textsc{DaiFu} can be readily used in the development of DL systems today to improve development productivity and save computing resources.



\end{document}